\documentclass[11pt]{revtex4}
\usepackage{amsfonts}
\usepackage{amsmath}
\usepackage{amssymb,epsf}
\usepackage{subfigure}
\usepackage{amssymb}
\usepackage{latexsym}
\usepackage{epsfig}            
\usepackage{float}
\usepackage{graphicx}

\usepackage{subfigure}
\begin{document}

\title{Effects of Backreaction on Power-Maxwell Holographic Superconductors in
Gauss-Bonnet Gravity}
\author{Hamid Reza Salahi$^{1}$\footnote{hrsalahi@shirazu.ac.ir},
Ahmad Sheykhi$^{1,2}$\footnote{asheykhi@shirazu.ac.ir} and Afshin
Montakhab $^{1}$ \footnote{montakhab@shirazu.ac.ir}}
\address{$^1$ Physics Department and Biruni Observatory, College of
Sciences, Shiraz University, Shiraz 71454, Iran\\
$^2$ Research Institute for Astronomy and Astrophysics of Maragha
(RIAAM), P.O. Box 55134-441, Maragha, Iran}

\begin{abstract}
We analytically and numerically investigate the properties of
$s$-wave holographic superconductors by considering the effects of
scalar and gauge fields on the background geometry in five
dimensional Einstein-Gauss-Bonnet gravity. We assume the gauge
field to be in the form of the Power-Maxwell nonlinear
electrodynamics. We employ the Sturm-Liouville eigenvalue problem
for analytical calculation of the critical temperature and the
shooting method for the numerical investigation. Our numerical and
analytical results indicate that higher curvature corrections
affect condensation of the holographic superconductors with
backreaction. We observe that the backreaction can decrease the
critical temperature of the holographic superconductors, while the
Power-Maxwell electrodynamics and Gauss-Bonnet coefficient term
may increase the critical temperature of the holographic
superconductors. We find that the critical exponent has the
mean-field value $\beta=1/2$, regardless of the values of Gauss-Bonnet
coefficient, backreaction and Power-Maxwell parameters.
\end{abstract}

\maketitle
\section{Introduction}
In $2008$, Hartnol et.al., put forwarded a new step on the
application of the gauge/gravity duality in condensed-matter
physics \cite{Har,Har2}. They have claimed that some properties of
strongly coupled superconductors can be potentially described by
classical general relativity living in one higher dimension. This
novel idea is usually called \textit{holographic superconductors}.
The motivation is to shed light on the understanding the mechanism
governing the high-temperature superconductors in condensed-matter
physics. The holographic $s$-wave superconductor model known as
Abelian-Higgs model was first established in \cite{Har,Har2}. The
well-known duality between anti-de Sitter (AdS) spacetime and the
conformal field theories (CFT) \cite{Mal} implies that there is a
correspondence between the gravity in the $d$-dimensional
spacetime and the gauge field theory livening on its
$(d-1)$-dimensional boundary. According to the idea of the
holographic superconductors given in \cite{Har}, in the gravity
side, a Maxwell field and a charged scalar field are introduced to
describe the $U(1)$ symmetry and the scalar operator in the dual
field theory, respectively. This holographic model undergoes a
phase transition from black hole with no hair (normal
phase/conductor phase) to the case with scalar hair at low
temperatures (superconducting phase) \cite{Gub}.

Following \cite{Har,Har2}, an overwhelming number of papers have appeared which try to investigate various properties of the
holographic superconductors from different perspective
\cite{Hor,Mus,RGC1,P.GWWY, P.BGRL, P.MRM, P.CW, P.ZGJZ,RGC2}. The
studies were also generalized to other gravity theories. In the
context of Gauss-Bonnet gravity, the phase transition of the
holographic superconductors were explored in
\cite{Wang1,Wang2,RGC3,Ruth, GBHSC}. The motivation is to study
the effects of higher order gravity corrections on the critical
temperature of the holographic superconductors. Considering the
holographic $p$-wave and $s$-wave superconductors in
$(3+1)$-dimensional boundary field theories, it was shown that
when Gauss-Bonnet coefficients become larger the operators on the
boundary field theory will be harder to condense \cite{RGC3}.
Taking the backreaction of the gauge and scalar field on the
background geometry into account, numerical as well as analytical
study on the holographic superconductors in five dimensional
Einstein-Gauss-Bonnet gravity were carried out in \cite{Ruth}. It
was observed that the temperature of the superconductor decreases
with increasing the backreaction, although the effect of the
Gauss-Bonnet coupling is more subtle: the critical temperature
first decreases then increases as the coupling tends towards the
Chern-Simons value in a backreaction dependent fashion
\cite{Ruth}.

In addition to the correction on the gravity side of the action,
it is also interesting to consider the corrections to the gauge
field on the matter side of the action. In particular, it is
interesting to investigate the effects of the nonlinear
corrections to the gauge field on the condensation and critical
temperature of the holographic superconductors. It was argued that
in the Schwarzschild AdS black hole background, the higher
nonlinear electrodynamics corrections make the condensation harder
\cite{Zi,shey1}. When the gauge field is in the form of
Born-Infeld nonlinear electrodynamics, analytical study, based on
the Sturm-Liouville eigenvalue problem, of holographic
superconductors in Einstein \cite{AnalyBI} and Gauss-Bonnet
gravity \cite{Gan1,Lala} have been carried out. In the background
of $d$-dimensional Schwarzschild AdS black hole, the properties of
Power-Maxwell holographic superconductors have been explored in
the probe limit \cite{PM} and away from the probe limit
\cite{PMb}. In our recent paper \cite{SSM}, we have analytically
as well as a numerically studied the holographic $s$-wave
superconductors in Gauss-Bonnet gravity with Power-Maxwell
electrodynamics. However, in that work, we did not investigate
the effects of backreaction and limited our study to the case
where scalar and gauge fields do not have an effect on the
background metric. Our purpose in the present work is to disclose
the effects of the backreaction on the phase transition and
critical temperature of the Power-Maxwell holographic
superconductors in Gauss-Bonnet gravity.

The organization of this paper is as follows. In the next section,
we provide the basic field equations of  Power-Maxwell holographic
superconductors in the background of Gauss-Bonnet-AdS black holes
by taking into account the backreaction. In section \ref{M}, based
on the Sturm-Liouville eigenvalue problem, we find a relation
between the critical temperature and charge density of the
backreacting holographic superconductor with Maxwell field in
Gauss-Bonnet gravity. In section \ref{PM}, we extend the study to
the case of Power-Maxwell nonlinear electrodynamics. By applying
the shooting method, we also compare our analytical calculations
with numerical results in this section. In section \ref{CriE}, we
calculate the critical exponent and the condensation values of the
Power-Maxwell holographic superconductor with backreaction. We
finish with conclusion and discussion in section \ref{Con}.

\section{Backreacting Gauss-Bonnet Holographic Superconductors
}\label{Int} To study a $(3+1)$-dimensional holographic
superconductor, we begin with a $(4+1)$-dimensional action of
Einstein-Gauss-Bonnet-AdS gravity which is coupled to
Power-Maxwell field and a charged scalar field,
\begin{eqnarray}\label{Act}
S=&\int d^{5}x \sqrt{-g}\frac{1}{2\kappa^2} \left[( R-2 \Lambda)
+\frac{\alpha}{2}
(R^{2}-4 R^{\mu\nu}
R_{\mu\nu}+R^{\mu\nu\rho\sigma}R_{\mu\nu\rho\sigma}\right]
\nonumber \\&+
\int d^{5}x \sqrt{-g}\left[-b(F_{\mu\nu}F^{\mu\nu})^q-
|\nabla\psi- ieA \psi|^2 - m^2 |\psi|^2 \right],
\end{eqnarray}
where $\kappa^2=8\pi G_5$ with $G_5$ is the $5$-dimensional
gravitational constant, $\Lambda =-{6}/{l^2}$ is the negative
cosmological constant, where $l$ is the AdS radius of spacetime,
and $\alpha$ is the Gauss-Bonnet coefficient. Here, $R$ and
$R_{\mu\nu}$ and $R_{\mu\nu\sigma\rho}$ are, respectively, Ricci
scalar, Ricci tensor and Riemann curvature tensor. $F^{\mu\nu}$ is
the electromagnetic field tensor and $q$ is the power parameter of
the Power-Maxwell field. $\psi$ is complex scalar field with the
charge $e$ and the mass $m$, and $A$ is the gauge field. Also, $b$
is coupling constant and due to positivity of energy density has
sign $(-1)^{q+1}$ \cite{HM,HM2}. For latter convenience we shall
take $b={(-1/2)^{q+1}}$. With this choice, the Power-Maxwell
Lagrangian will reduce to the Maxwell Lagrangian in the limit
$q=1$.

It is easy to check that by re-scaling $\psi \rightarrow
\tilde{\psi}/e$, $\phi \rightarrow \tilde{\phi}/e$ and $b
\rightarrow \tilde{b} e^{2q-2}$, a factor $1/e^2$ will appear
in front of matter part of action (\ref{Act}). Thus, the probe
limit can be deduced  when  $\kappa^2/e^2 \rightarrow 0$. In order
to take the backreaction into account, in this paper, we keep
$\kappa^2/e^2$ finite and for simplicity we set $e$ as unity.

Taking the backreaction into account, the plane-symmetric black
hole solution with an asymptotically AdS behavior in
$5$-dimensional spacetime may be written
\begin{eqnarray}\label{metric}
ds^{2}&=-e^{-\chi(r)} f(r)
dt^{2}+\frac{dr^2}{f(r)}+\frac{r^{2}}{l_{\rm
eff}^2}\left(dx^2+dy^2+dz^2\right),
\end{eqnarray}
where
\begin{eqnarray}\label{le}
l_{\rm eff}^2\equiv\frac{2
\alpha}{1-\sqrt{1-\frac{4\alpha}{l^2}}},
\end{eqnarray}
is the effective AdS radius of the spacetime. The ratio of $l_{\rm
eff}/l$ can be smaller than unity for $\alpha>0$, while for
$\alpha<0$ it is obvious that $l_{\rm eff}/l$ is larger than
unity.

Superconductivity phase transition is dual to formation of
charged matter field in the bulk, and for occurrence this phase
transition in bulk, one needs to prevent the charged matter field
to falls into the black hole, thus we expect greater curvature of
spacetime in bulk make condensation harder which corresponds to
the positive values of $\alpha$. Also for $\alpha<0$, we shall see
that the scalar field can be formed easier, means at higher
temperature.

The Hawking temperature of black hole is given by
\begin{eqnarray}\label{Temp}
T= \frac{f'(r_+) e^{-\chi(r+)/2}}{4\pi},
\end{eqnarray}
where $r_+$ is the black hole horizon and the prime denotes
derivative with respect to $r$. We choose the electromagnetic
gauge potential and scalar field as
\begin{eqnarray}\label{phipsi}
\psi=\psi(r), \ \ \ \ \ \ \ \ \  A=\phi(r)dt.
\end{eqnarray}
Without lose of generality, we can take $\phi(r)$ and $\psi(r)$
real. The equation of motions can be obtained by varying action
(\ref{Act}) with respect to the metric and matter fields. We find:
 \begin{eqnarray}\label{eompsi}
 \psi ''+ \psi '\left(\frac{f'}{f}+\frac{3}{r}-\frac{\chi'}{2}\right) +\psi
 \left(\frac{\phi^2}{f^2}-\frac{m^2}{f}\right)=0,
\end{eqnarray}
\begin{eqnarray}\label{eomphi}
\phi''+\phi' \left(\frac{3}{(2q-1)r}+\frac{\chi'}{2} \right)-
\frac{2e^{(1-q)\chi}\psi^2 \phi'^{2-2q}}{q(2q-1)f}\phi=0,
\end{eqnarray}
\begin{eqnarray}\label{eomchi}
 \chi' \Big(1-\frac{2\alpha f}{r^2} \Big) +
\frac{4r\kappa^2}{3} \Big(\psi'^2+
\frac{e^{\chi}\phi^2\psi^2}{f^2} \Big)=0,
\end{eqnarray}
\begin{eqnarray}\label{eomf}
 f' \Big(1-\frac{2\alpha f}{r^2} \Big) +\frac{2f}{r}-\frac{4r}{l^2}+
\frac{2r\kappa^2}{3} \Big(m^2 \psi^2+f\psi'^2+
\frac{e^{\chi}\phi^2\psi^2}{f}+\frac{(2q-1)}{2}
e^{q\chi}\phi'^{2q} \Big)=0.
\end{eqnarray}
In order to solve the above field equations, we need
appropriate boundary conditions both on the horizon $r_+$, which
is defined by $f(r_+)$=0, and on the AdS boundary where
$r\rightarrow \infty$. On the horizon, the regularity condition
imposes
\begin{eqnarray}\label{HBC1}
\phi(r_+)=0, \ \ \ \ \ \ \ \ \ \ \psi'(r_+)=\frac{m^2
\psi(r_+)}{f'(r_+)},
\end{eqnarray}
and thus from Eqs. (\ref{eomchi}) and (\ref{eomf}) we have
\begin{eqnarray}\label{Hchi}
 \chi'(r_+)=-\frac{4\kappa^2 r_+}{3}
\left( \psi'(r_+)^2+\frac{e^{\chi(r_+)}\phi'(r_+)^2 \psi(r_+)^2}
{f'(r_+)^2} \right),
\end{eqnarray}
\begin{eqnarray}\label{Hf}
 f'(r_+)= \frac{4r_+}{l^2} -\frac{2 \kappa^2 r_+}{3} \left(m^2 \psi(r_+)^2 +
\frac{(2q-1)}{2} e^{q\chi(r_+)}\phi'(r_+)^{2q} \right).
\end{eqnarray}
Since our solutions are asymptotically AdS, thus as $r\rightarrow
\infty$, we have
\begin{eqnarray}\label{BC}
\chi(r) \rightarrow 0, \ \  \ \ f(r) \approx \frac{r^2}{l_{\rm
eff}^2},
 \ \  \ \ \phi(r) \approx
\mu-\frac{\rho^{\frac{1}{2q-1}}}{r^{\frac{4-2q}{2q-1}}}, \ \  \ \
\psi \approx  \frac{\psi_{-}}{r^{\Delta_{-}}}  +\frac{\psi_{+}}
{r^{\Delta_{+}}},
\end{eqnarray}
where $\mu$ and $\rho$ are, respectively, chemical potential and
charge density of the CFT boundary, and $\Delta_{\pm}$ is defined
as
\begin{eqnarray}\label{delta}
\Delta_{\pm} = 2 \pm  \sqrt {4 + m^2 l_{\rm eff}^2}.
\end{eqnarray}
According to the AdS/CFT correspondence,
$\psi_{\pm}=<\mathcal{O_{\pm}}>$, where $\mathcal{O_{\pm}}$ is the
dual operator to the scalar field with the conformal dimension
$\Delta_{\pm}$. We have the freedom to impose boundary conditions
such that either $\psi_-$ or $\psi_+$ vanish. We prefer to keep fixed $\Delta_{\pm}$ while we vary $\alpha$, thus we
set $\tilde{m}^2= m^2 l_{\rm eff}^2$. For example, for
$\tilde{m}^2=-3$, we have $\Delta_+=3$ for all values of parameter
$\alpha$.

It is important to note that, unlike other known electrodynamics,
the boundary condition for the gauge field $\phi(r)$ given in Eq.
(\ref{BC}), depends on the power parameter $q$. Using boundary
condition (\ref{BC}) and the fact that $\phi$ should be finite as
$r \rightarrow \infty$, we require that $(4-2q)/{(2q-1)}>0$ which
restricts $q$ to ranges as $1/2<q<2$.

It is easier to work in the dimensionless variable, $z=r_+/r$,
instead of variable $r$. Under this transformation, equations of
motion (\ref{eompsi})-(\ref{eomf}) become
\begin{eqnarray}\label{eompsiz}
\psi'' + \left(\frac{f'}{f}-\frac{1}{z}-\frac{\chi'}{2}\right)\psi' +
\frac{r_{+}^2}{z^4}\left(\frac{\phi^2 e^{\chi}}{f^2}-
\frac{m^2}{f}\right)\psi=0,
\end{eqnarray}
\begin{eqnarray}\label{eomphiz}
\phi'' + \left(\frac{4q-5}{(2q-1)z}+\frac{\chi'}{2}\right)\phi' -
\frac{2r_{+}^{2q}\psi^2 \phi'^{2-2q}}{(-1)^{2q} q
(2q-1) z^{4q} f}\phi=0,
\end{eqnarray}
\begin{eqnarray}\label{chiz}
\chi' \left(1-\frac{2 \alpha  z^2 f}{r_+^2}\right)-\frac{4 \kappa
^2 r_+^2}{3 z^3} \left(\frac{e^{\chi } \psi^2
\phi^2}{f^2}+\frac{z^4 \psi '^2}{r_+^2}\right)=0,
\end{eqnarray}
\begin{eqnarray}\label{fz}
&& f' \left(1-\frac{2 \alpha  z^2 f}{r_+^2}\right)-\frac{2
f}{z}+\frac{4 r_+^2}{ l^2z^3}-\frac{2\kappa ^2 r_+^2}{3 z^3}
\Bigg[\frac{z^4 f \psi'^2}{r_+^2} \nonumber
\\ && +\frac{e^{\chi} \psi^2 \phi^2}{f}
 +m^2 \psi^2 -\frac{1}{2} (1-2 q) e^{q \chi} (-1)^{2 q}\left(\frac{z^2 \phi'}{r_+}\right)^{2
 q}\Bigg]=0.
\end{eqnarray}
Here the prime indicates the derivative with respect to the new
coordinate $z$ which ranges in the interval $[0,1]$, where
$z=0$ and $z=1$ correspond to the boundary and horizon,
respectively. Since near the critical point the expectation value of
scalar operator ($<\mathcal{O_{\pm}}>$) is small, we can select
it as an expansion parameter
\begin{eqnarray}\label{so}
\epsilon \equiv <\mathcal{O}_i>,
\end{eqnarray}
where $i=\pm$. Using the fact that $\epsilon \ll$, we can expand $f$ and $\chi$ around
the Gauss-Bonnet AdS spacetime as
\begin{eqnarray}\label{fex}
f=f_{0}+\epsilon^2 f_{2}+\epsilon^4f_{4}+...,
\end{eqnarray}
\begin{eqnarray}\label{chiexpand}
\chi=\epsilon^2\chi_{2}+\epsilon^4\chi_{4}+....
\end{eqnarray}
Note that since we are interested in solution in which condensation is small, $\psi$ and
$\phi$
can also be expanded as
\begin{eqnarray}
\psi=\epsilon \psi_{1}+\epsilon^3 \psi_{3}+\epsilon^5\psi_{5}+...,
\end{eqnarray}
\begin{eqnarray}\label{phiex}
\phi=\phi_{0}+\epsilon^2\phi_{2}+\epsilon^4\phi_{4}+...
\end{eqnarray}
We further assume the chemical potential is expanded as
\cite{CPH},
\begin{eqnarray}
\mu=\mu_{0}+\epsilon^2\delta \mu_{2}+...,
\end{eqnarray}
where $\delta \mu_{2}>0$. Thus near the critical point for the
order parameter as the function of chemical potential we have
\begin{eqnarray}
\epsilon\thickapprox\Bigg(\frac{\mu-\mu_{0}}{\delta
\mu_{2}}\Bigg)^{1/2},
\end{eqnarray}
It is obvious when  $\mu \rightarrow \mu_{0}$, the order parameter
approaches zero which indicate phase transition point. Thus phase
transition occurs at the critical value $\mu_{c}=\mu_{0}$. Let us
note that the order parameter grows with exponent $1/2$ which is
the universal result from the Ginzburg-Landau mean field theory.

In the next two sections we solve the field equations
(\ref{eompsiz})-(\ref{fz}) by using expansions
(\ref{fex})-(\ref{phiex}), for the linear Maxwell field  as well
as the nonlinear Power-Maxwell electrodynamics.
\section{Critical temperature of GB holographic superconductors with Maxwell
field}\label{M} In this section, by using the Sturm-Liouville
eigenvalue problem, we obtain the relation between
the critical temperature and charge density of the $s$-wave
holographic superconductor with backreaction in Gauss-Bonnet-AdS
black holes. The Maxwell theory corresponds to $q=1$.

Employing the matching method, the holographic superconductors in
Gauss-Bonnet gravity with backreaction for the Maxwell \cite{SK}
and the nonlinear Born-Infeld electrodynamics have been studied
\cite{Gan1}. However, it was shown that the matching method is less
accurate than Sturm-Liouville method and the obtained results from
Sturm-Liouville method are in a better agreement with the
numerical results.

At zeroth order for the expansion parameter, Eq.
(\ref{eomphiz}) may be written as
\begin{eqnarray}
\phi_0''(z) - \frac{\phi_0'(z)}{z} =0,
\end{eqnarray}
which is the equation of motion of the electromagnetic field in
the Maxwell theory and has solution $\phi_0(z)=\mu_0 \left(1-z^2
\right)$ with $\mu_0 = \rho/r_+^2$. At the critical point, we have
$\mu_0=\mu_c= \rho/r_{+c}^2$, where $r_{+c}$ is the radius of the
horizon at the phase transition point. Therefore, solution of
$\phi_0 (z)$ at the critical point may be written as
\begin{eqnarray}\label{phi0}
\phi_0=r_{+c} \zeta \left(1-z^2 \right), \ \ \ \ \ \ \ \   \zeta
\equiv \rho/r_{+c}^3.
\end{eqnarray}
Inserting back this solution into Eq. (\ref{fz}), we find the
metric function at the zeroth order:
\begin{eqnarray}\label{f0}
f_0(z)=r_+^2 g(z)=\frac{r_+^2}{2 \alpha z^2}
\left(1-\sqrt{1-\frac{4 \alpha}{l^2}  \left(1-z^4\right) +\frac{8
\alpha}{3} \zeta ^2 \kappa ^2 z^4 \left(1-z^2\right)}\right),
\end{eqnarray}
where we have used the fact that on the horizon $f_{0}(1)=0$, and
we have defined a new function $g(z)$ for  convenience. We
note that $f_{0}(z)$ restores the metric function of
Gauss-Bonnet-AdS gravity in the probe limit as
$\kappa\rightarrow0$.

At the first order approximation, the asymptotic AdS boundary
conditions for $\psi$ can be expressed as
\begin{eqnarray}
\psi_{1} \approx \frac{\psi_{-}}{r_{+}^{\_{-}}}
z^{\Delta_{-}}+ \frac{\psi_{+}}{r_{+}^{\Delta_{+}}}z^{\Delta_{+}}.
\end{eqnarray}
Near the boundary $z=0$, we introduce trial function $F(z)$
\begin{eqnarray}\label{psiF}
\psi_{1}(z)= \frac{<\mathcal{O}_i>}{r_{+}^{\triangle_{i}}} z^ {\triangle_{i}}F(z),
\end{eqnarray}
with boundary condition $F(0)=1$ and $F'(0)=0$. Substituting Eq.
(\ref{psiF}) into (\ref{eompsiz}) we arrive at
\begin{eqnarray}\label{F}
&&F''(z)+F'(z) \left(\frac{g'(z)}{g(z)}+\frac{2 \Delta_i -1}{z}
\right)\nonumber
\\ &&+F(z)\Bigg[ \frac{\Delta_i}{z} \left( \frac{g'(z)}{g(z)}
+\frac{\Delta_i -2}{z}\right) \Delta_i z^2 g(z)^2 -\frac{m^2}{z^4
g(z)}+\frac{\zeta ^2 (z^2-1)^2}{g(z)^2 z^4}\Bigg]=0.
\end{eqnarray}
We can convert Eq. (\ref{F}) into the Standard Sturm-Liouville
equation, namely
\begin{eqnarray}\label{SL}
[T(z)F'(z)]' - Q(z) F(z)+\zeta^2 P(z) F(z)=0,
\end{eqnarray}
where
\begin{eqnarray}\label{QP}
 Q(z)&=&-T(z)\Bigg[ \frac{\Delta_i}{z} \left( \frac{g'(z)}{g(z)}+\frac{\Delta_i
-2}{z}\right) \Delta_i  z^2 g(z)^2 -\frac{m^2}{z^4 g(z)}\Bigg],
\nonumber \\  P(z)&=& T(z)\frac{ (z^2-1)^2}{g(z)^2 z^4}.
\end{eqnarray}
According to the Sturm-Liouville eigenvalue problem, $\zeta^2$ can
be obtained via
\begin{eqnarray}\label{zeta2}
\zeta^2=\frac{\int_{0}^{1}[T(z)[F'(z)]^2+Q(z)F^2(z)]dz}{\int_{0}^{1}P(z)F^2(z)dz}.
\end{eqnarray}
In order to determine $T(z)$ we need to solve equation
\begin{eqnarray}\label{Tz}
T(z) p(z)=T'(z),
\end{eqnarray}
where $p(z)$ is
\begin{eqnarray}\label{p}
p(z)=\left(\frac{g'(z)}{g(z)}+\frac{2 \Delta_i -1}{z} \right).
\end{eqnarray}
Since $\alpha$ is small, we can expand the above expression for
$p(z)$ and keep terms up to $\mathcal{O}(\alpha^2)$. Then we put
the result in Eq.~(\ref{Tz}) and obtain the following solution
for $T(z)$
\begin{eqnarray}
T(z)&=&z^{2 \Delta_i +1} \Bigg(3 \left(z^{-4}-1 \right)+2 \zeta ^2
\kappa ^2 \left(z^2-1\right)\Bigg)\nonumber
\\ && \times \exp\Bigg\{\Bigg(2+\alpha
\left[2 \zeta ^2 \kappa ^2 z^4(z^2-1)-3z^4+6\right]\Bigg)
\frac{\alpha  z^4}{6}  \Bigg(2 \zeta ^2 \kappa ^2
\left(z^2-1\right)-3\Bigg) \Bigg\}.
\end{eqnarray}
For small backreaction parameter, $\kappa$, the explicit
expressions for $T(z)$, $Q(z)$ and $P(z)$ up to second order terms
of $\alpha$ and $\kappa$, are given by
\begin{eqnarray}
T(z) &\approx & z^{2 \Delta_i +1} \Bigg \{ 3 (z^{-4} -1)+\Bigg[ 2
\zeta ^2 \kappa ^2 \Big(z^2-1\Big) \Bigg(1+\alpha  \Big(1+3
\alpha-2
(5 \alpha +1) z^4 +6 \alpha  z^8\Big) \nonumber \\
&&- \alpha  \Big(z^2+1\Big)
\Big(\alpha  \left(2 z^4-3\right)-1\Big) \Bigg) \Bigg] \Bigg\}
+\mathcal{O}(\alpha^3)+\mathcal{O}(\kappa^4),
\end{eqnarray}
\begin{eqnarray}
&& Q(z) \approx z^{2 \Delta_i -5} \Bigg\{ 3 \Delta_i  \Bigg(4+s
\Delta_i  z^4-\Delta_i +2 \alpha ^2 (\Delta_i +8) z^{12}-\alpha  (5 \alpha +1)
(\Delta_i +4) z^8\Bigg)\nonumber
 \\ && +2 \Delta_i  \zeta ^2 \kappa ^2 z^4
\Bigg(6 \alpha ^2 z^8 \Big[\Delta_i +8-(\Delta_i +10)
z^2\Big]
 -2 \alpha  (5 \alpha +1) z^4 \Big[\Delta_i +4-(\Delta_i +6)
z^2\Big] \nonumber \\
&&+s \Delta_i
 -s (\Delta_i +2) z^2+\Bigg)+3 \tilde{m}^2 \Bigg\}+\mathcal{O}(\alpha^3)+\mathcal{O}(\kappa^4),
\end{eqnarray}
\begin{eqnarray}
P(z) &&\approx  \frac{1}{(z^2+1)^2} z^{2 \Delta_i -3}  (z^2-1) \Bigg\{3
\left(\alpha ^2+2 \alpha -1\right) \left(z^2+1\right)-z^4 \left[2
(1-\alpha ) \zeta ^2 \kappa ^2+3 \alpha  (\alpha +1)\right]\nonumber
\\ &&  -3 \alpha  (\alpha +1) z^6+\alpha ^2 z^8 \left(4 \zeta ^2 \kappa
^2+3\right)+3 \alpha ^2 z^{10} -2 \alpha ^2 \zeta ^2 \kappa ^2 z^{12}
\Bigg \}+\mathcal{O}(\alpha^3)+\mathcal{O}(\kappa^4),
\end{eqnarray}
where $s=3 \alpha^2+\alpha+1$ and hereafter we set $l=1$ for
simplicity. In order to use Sturm-Liouville eigenvalue problem, we
will use iteration method in the rest of this section. We take
$\kappa=\kappa_n \Delta \kappa$ where $\Delta
\kappa=\kappa_{n+1}-\kappa_n$ is step size of iterative procedure
and we choose $\Delta \kappa=0.05$. Using the fact that
 \begin{eqnarray}
\zeta^2 \kappa^2 = \zeta^2
\kappa_n^2=\Big(\zeta^2|_{\kappa_{n-1}}\Big)\kappa_n^2+
\mathcal{O}(\Delta \kappa)^4,
 \end{eqnarray}
and taking $\kappa_{-1}=\zeta|_{\kappa_{-1}}=0$, we obtain the
minimum eigenvalue of Eq.~(\ref{SL}). We also take the trial function
$F(z)=1-a z^2$. For example for $\tilde{m}^2=-3$, $\alpha=0.05$
and $\kappa=0$, we have
\begin{eqnarray}
\zeta^2_{\kappa_0}=\frac{-566.794 a^2+1096.44 a-737.301}{-7.02708
a^2+24.3982 a-26.0408},
\end{eqnarray}
which attains its minimum $\zeta^2_{\rm min}=19.9456$ for
 $a=0.7147$. In the second iteration, we take $\kappa=0.05$ and
 $\zeta^2|_{\kappa_0}=19.9456$ in calculation of integrals in Eq.~(\ref{zeta2}), and
therefore for $\zeta^2_{\kappa_1}$, we get
 \begin{eqnarray}
  \zeta^2_{\kappa_1}=\frac{-559.863 a^2+1083.88 a-730.968}{-7.09007 a^2+24.5832
  a-26.189},
  \end{eqnarray}
 which has the minimum value $\zeta^2_{\rm min}=19.7936$ at  $a=0.7119$.
In the Table \ref{tab1} we summarize our results for $\zeta_{\rm
min}$ and $a$ with different values of Gauss-Bonnet coupling
parameter $\alpha$, backreaction parameter $\kappa$ and reduced
mass of scalar field $\tilde{m}^2$.
\begin{table}[ht]
\begin{center}
\begin{tabular}{|c|c|c|c|c|c|c|c|c|}
\hline &\multicolumn{2}{c|}{$\kappa=0$}
&\multicolumn{2}{c|}{$\kappa=0.05$} &\multicolumn{2}{c|} {$
\kappa=0.10$}
&\multicolumn{2}{c|}{$\kappa=0.15$} \\
\hline
$\alpha$ &$a$ & $\zeta_{\rm min}^2$ & $a $ &$\zeta_{\rm min}^2$ & $a$ & $\zeta_{\rm min}^2$  & $a$ & $\zeta_{\rm min}^2$ \\
\hline
$-0.19$ &0.7344 &14.0472   &0.7330 &13.9836 &0.7287 &13.7949&0.7213 &13.4909 \\
\hline
$-0.1$ &0.7307&15.693 &0.7290 &15.6105 &0.7238&15.3662&0.7146&14.9745\\
\hline
$0$  &0.7218& 18.2300&0.7195 &18.1097 &0.7123&17.7546&0.6996&17.1902\\
\hline
$0.1$ &0.7050&22.1278 &0.7015 &21.9279 &0.6904&21.3407&0.6705&20.4209\\
\hline
$0.2$ &0.67304&28.9837 &0.6667 &28.5719 &0.6462&27.3751&0.6081&25.5561\\
\hline
\end{tabular}
\caption{Analytical results of $\zeta^2_{\rm min}$ and $a$ for
Maxwell case with different values of the backreaction $\kappa$
and GB parameter $\alpha$ for $\lambda_+$. Here we have taken
$\tilde{m}^2=-3$.}\label{tab1}
\end{center}
\end{table}
\begin{center}
\begin{table}[ht]
\begin{tabular}{|c|c|c|c|c|c|c|c|c|}
\hline &\multicolumn{2}{c|}{$\kappa=0.05$} &\multicolumn{2}{c|} {$
\kappa=0.10$}
&\multicolumn{2}{c|}{$\kappa=0.15$} \\
\hline $\alpha$ & Analytical & Numerical &Analytical & Numerical
&Analytical & Numerical  \\
\hline $-0.19$   & 0.2027 $\rho^{1/3}$ & 0.2050 $\rho^{1/3}$
&0.1961 $\rho^{1/3}$ & 0.1986 $\rho^{1/3}$ & 0.1854
$\rho^{1/3}$&0.1882
$\rho^{1/3}$ \\
\hline $-0.1$  &0.1987 $\rho^{1/3}$&0.2008 $\rho^{1/3}$&0.1915
$\rho^{1/3}$&0.1938
$\rho^{1/3}$&0.1800 $\rho^{1/3}$&0.1825 $\rho^{1/3}$ \\
\hline $0$  &0.1935 $\rho^{1/3}$&0.1953 $\rho^{1/3}$&0.1854
$\rho^{1/3}$&0.1874 $\rho^{1/3}$&0.1726 $\rho^{1/3}$&0.1764 $\rho^{1/3}$\\
\hline $0.1$  &0.1868 $\rho^{1/3}$&0.1882 $\rho^{1/3}$&0.1775
$\rho^{1/3}$&0.1791 $\rho^{1/3}$&0.1630 $\rho^{1/3}$&0.1646 $\rho^{1/3}$\\
\hline $0.2$  &0.1771 $\rho^{1/3}$&0.1779 $\rho^{1/3}$&0.1666
$\rho^{1/3}$&0.1668 $\rho^{1/3}$&0.1499 $\rho^{1/3}$&0.1500 $\rho^{1/3}$\\
\hline
\end{tabular}
\caption{Comparison of analytical and numerical values of the
critical temperature for Maxwell case with
$\tilde{m}^2=-3$.}\label{tab2}
\end{table}
\end{center}
Combining Eqs. (\ref{Temp}), (\ref{Hf}), (\ref{phi0}) and using
definition of $\zeta$, we obtain the following expression for the
critical temperature
\begin{eqnarray}\label{Tc}
T_c= \frac{1}{ \pi} \left(1-\frac{ \kappa^2 \zeta^2_{\rm
min}}{3}\right) \left[\frac{\rho}{\zeta_{\rm min}}\right]^{1/3}.
\end{eqnarray}
We apply the iterative procedure to obtain critical temperature
for different values of $\alpha$, $\kappa$ and $\tilde{m}^2$. In
table \ref{tab2} we summarize critical temperature of phase
transition of holographic superconductor in Maxwell
electrodynamics for $\Delta_+$ obtained analytically from
Sturm-Liouville method. For comparison, we also provide numerical
results which we obtain by using shooting method. In this
numerical method we solve Eq. (\ref{eompsiz}) with $\phi(z)$ and
$f(z)$  given in Eqs. (\ref{phi0}) and  (\ref{f0}). Then we find
the critical charge density $\rho$ which satisfy the boundary
condition $\psi_-=0$ in $z \rightarrow 0$. We obtain discrete
values of critical $\rho$ which had this situation. Due to the
stability condition \cite{SSGubser}, we chose the lowest value of
$\rho_c$ and by using dimensionless quantity $T^3/\rho$ we
calculated critical temperature of the phase transition for
different values of Gauss-Bonnet parameter and backreaction
parameter.
\section{Critical temperature of GB holographic superconductor with
Power-Maxwell field} \label{PM} In this section we investigate the
behavior of holographic superconductor for the general case $q
\neq 1$ away from probe limit in the Gauss-Bonnet gravity. Just
like previous section, we need solution of Eqs. (\ref{eomphiz}),
(\ref{chiz})  and (\ref{fz}) in order to solve (\ref{eompsiz}).
Using expansion (\ref{fex})-(\ref{phiex}) and at the zeroth order
of small parameter $\epsilon$, one can easily check that $\phi_0$
and $g$ have the following solution
\begin{eqnarray}\label{phiPM}
\phi_0(z)=\zeta  r_{+_c} \left(1-z^{\frac{2 (2-q)}{2 q-1}}\right), \ \ \ \ \ \zeta=\frac{\rho^{\frac{1}{2q-1}}}{r_{+_c}^{\frac{3}{2q-1}}},
\end{eqnarray}
\begin{eqnarray}\label{gz}
g(z)=\frac{1}{2 \alpha  z^2} \left(1-\sqrt{1-4 \alpha
\left(1-z^4\right)-\frac{4^q (2-q)^{2
q-1}}{3 (2 q-1)^{2 q-2}} 2\alpha  \kappa ^2 \zeta ^{2 q} \left[\left(z^{\frac{6 q}{2 q-1}}\right)-z^4\right]}\right),\nonumber \\
\end{eqnarray}
where
$g(z)=f_0(z)/r_+^2$. Expanding the above expression for $g(z)$ up
to $\mathcal{O} (\kappa^4)$ and $\mathcal{O} (\alpha^2)$, one gets
\begin{eqnarray} \label{gz}
 g(z) && \approx \frac{1}{z^2}-z^2+\frac{\left(z^4-1\right)^2}{z^2}
\left[\alpha -2 \alpha ^2 \left(z^4-1\right)\right]+\frac{ (2
q-1)^{2-2
q} (4-2 q)^{2 q-1} \left(z^{\frac{6 q}{2 q-1}}-z^4\right)}{3 z^2} \nonumber \\
&& \times \Big\{[6\alpha ^2 \left(z^4-1\right)^2 -2 \alpha
\left(z^4-1\right)+1\Big\}\kappa ^2 \zeta ^{2 q}-\frac{(2
q-1)^{4-4 q} (4-2 q)^{4 q-2}
\left(z^4-z^{\frac{6 q}{2 q-1}}\right)^2}{9 z^2}  \nonumber \\
&&\times \left[ 6 \alpha ^2
\left(z^4-1\right)-\alpha\right] \kappa ^4 \zeta ^{4 q}+\mathcal{O}(\alpha^3)+\mathcal{O}(\kappa^6).
\end{eqnarray}
One may substitute Eq.~(\ref{psiF}) and Eq.~(\ref{phiPM}) into
Eq.~(\ref{eompsiz}) and get an expression for $F(z)$, and then
converting it to the Sturm-Liouville equation form (\ref{SL}), resulting in:
\begin{eqnarray} \label{TTz}
&& T(z) \approx z^{2 \Delta -3} \Bigg\{1-z^4 \Big[ \alpha (z^4-1)
\Big(\alpha  \left(2 z^4-3\right)-1\Big)\Big]+\frac{ (4-2 q)^{2
q} (2 q-1)^{2-2 q}}{6(q-2)} \nonumber \\
&&\times \Bigg[1-z^{\frac{6 q}{2
q-1}} \Big(1+\alpha +\alpha ^2 \left(6 z^8+3\right)\Big) -\alpha  z^8 \left(1+5 \alpha -4
\alpha  z^4-2 (5 \alpha +1) z^{\frac{4-2q}{2 q-1}}\right)\Bigg]\kappa
^2\zeta ^{2 q}\nonumber \\
&& + \Bigg[1+\alpha  (5 \alpha +1) z^{\frac{12 q}{2
q-1}}+z^{\frac{6 q}{2 q-1}} \Big[\alpha  \Big(\alpha  \left(6
z^8-3\right)-1\Big)-1\Big]-2 \alpha ^2 z^{10} \left(3 z^{\frac{6}{2
q-1}}+z^2\right)\Bigg] \kappa ^4\zeta ^{4 q} \nonumber \\
&& \times\frac{2^{4 q} (2-q)^{4 q-2} (1-2 q)^{4-4 q}}{36}\Bigg\},
\\
&&P(z)\approx T(z) \frac{\left(1-z^{\frac{2 (2-q)}{2
q-1}}\right)^2}{z^4 g(z)^2}+\mathcal{O}(\alpha^4)
+\mathcal{O}(\kappa^6),\\
&&Q(z)\approx-T(z)\Bigg\{ \frac{\Delta_i}{z} \left(
\frac{g'(z)}{g(z)}+\frac{\Delta_i -2}{z}\right) \Delta_i  z^2
g(z)^2 -\frac{m^2}{z^4 g(z)}\Bigg\}+\mathcal{O}(\alpha^4)
+\mathcal{O}(\kappa^6).
\end{eqnarray}
Again, using Eq.~(\ref{SL}), with trial function $F(z)=1-az^2$, we
obtain the minimum eigenvalue $\zeta^2_{\rm min}$ for the
Power-Maxwell electrodynamic case. For example, with $q=3/4$,
$\alpha=0.1$, $\kappa=0.05$, and $\tilde{m}^2=-3$ and using
iterative procedure, we get
\begin{eqnarray}\label{zetamin}
\zeta_{\kappa_1}^2=\frac{30 \left(1.0333 a^2-2.0068
a+1.3652\right)}{1.3539 a^2-4.1680 a+3.6937}.
\end{eqnarray}
Varying $\zeta_{\kappa_1}^2$ with respect to $a$ to find minimum
value of $\zeta^2$, we obtain $\zeta_{\rm min}^2=9.50679$ at
$a=0.5675$. Also for the case $q=5/4$, $\alpha=-0.19$,
$\kappa=0.1$ and $\tilde{m}^2 =0$ we obtain
\begin{eqnarray}\label{zetamin2}
\zeta_{\kappa_2}^2=\frac{461.3339 a^2-968.8766
a+593.0286}{a^2-3.2657 a+3.0859},
\end{eqnarray}
which attains its minimum $\zeta_{\rm min}=98.9682$ at $a=0.8909$.
Then, we find the critical temperature from Eqs. (\ref{Temp}),
(\ref{Hf}) and (\ref{phiPM}) as
\begin{eqnarray}\label{TPM}
T_c= \frac{1}{4\pi} \Bigg[4-\frac{(4-2q)^{2q}}{3
(2q-1)^{2q-1}}\kappa^2 \zeta_{\rm
min}^{2q}\Bigg]\Big(\frac{\rho}{\zeta_{\rm
min}^{2q-1}}\Big)^{\frac{1} 3}.
\end{eqnarray}
Clearly, $T_{c}$ depends on the Power-Maxwell parameter $q$,
Gauss-Bonnet parameter $\alpha$ and backreaction parameter
$\kappa$.
\begin{figure}
\centering
\subfigure[$\kappa=0.05$]{\includegraphics[scale=.85]{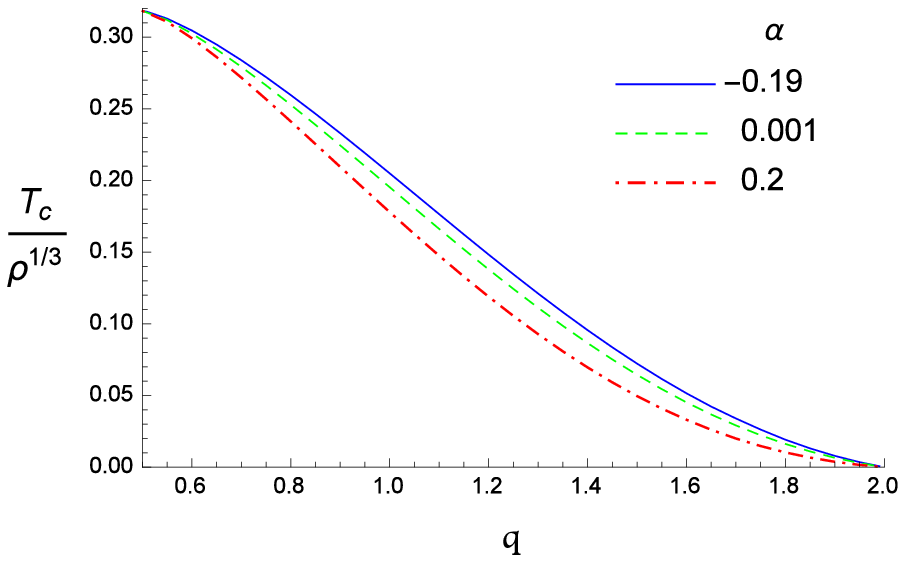}\label{fig1a}}\hspace*{.2cm}
\subfigure[$\alpha=0.10$]{\includegraphics[scale=.85]{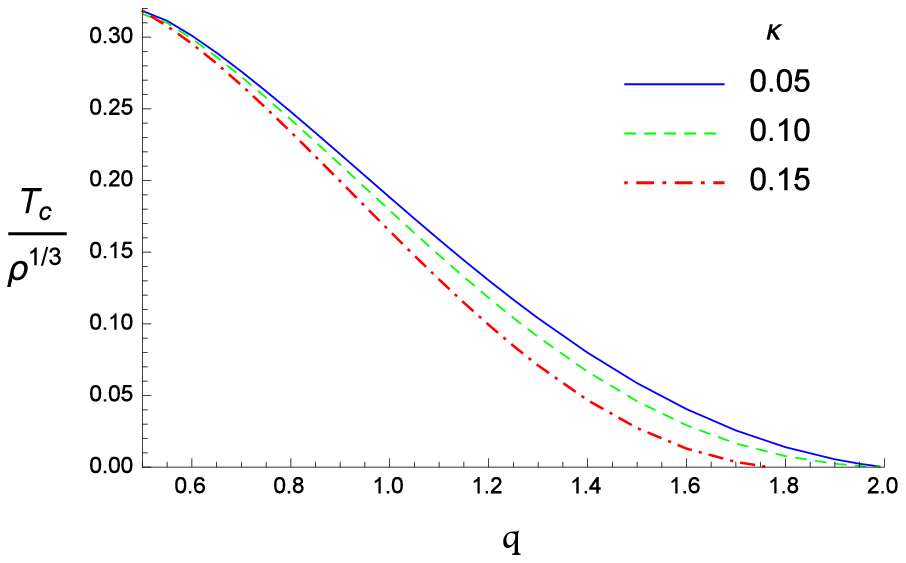}\label{fig1b}}\hspace*{.2cm}
\caption{Critical temperature of GB holographic superconducting
phase transition with Power-Maxwell field as a function of $q$ for
$\tilde{m}^2=-3$.}\label{fig1}
\end{figure}
In Fig. \ref{fig1}, we present reduced critical temperature of
phase transition for a $(3+1)$-dimensional holographic
superconductor as a function of $q$ with different values of
$\kappa$ and $\alpha$. For simplicity, we focus on the boundary
condition which $\psi_-=0$,  and as an example, we take
$\tilde{m}^2=-3$ in these figure.

In Fig. \ref{fig1}(a) we fix the backreaction parameter to
$\kappa=0.05$ in order to investigate behavior of critical
temperature as a function of power parameter $q$ for three allowed
value of Gauss-Bonnet parameter. It clearly indicates that for any
values of $\alpha$, by decreasing $q$, superconductor phase is
more accessible. Also, we find out that in the presence of
backreaction of the matter fields on the metric, increasing
Gauss-Bonnet parameter $\alpha$ makes condensation harder and and
thus the critical temperature of the phase transition decreases.
It is interesting that decreasing $\alpha$ from zero to negative
values in the allowed range can cause the phase transition to
superconductor phase easier for any values of the power parameter
$q$.

We also provide Fig. \ref{fig1}(b) by fixing the Gauss-Bonnet
parameter to $\alpha=0.1$ for studying the behavior of reduced
critical temperature in terms of the power parameter $q$ for
different values of the backreaction parameter $\kappa$. From this
figure we see that for any values of $q$, by increasing the
backreaction of the matter fields on the background geometry,
which is corresponding to decreasing the charge of the scalar
field,  the phase transition is made harder in the
Einstein-Gauss-Bonnet gravity.

We mention that in the allowed range of the power parameter, there
exist some un-physical regimes in which critical temperature
becomes negative. For example, by increasing backreaction
parameter to greater values, we may obtain negative $T_c$ which
means for some values of the power parameter we do not have phase
transition if complex field charge is less than some critical
charge. Here we disregard these regimes and work in regimes with
positive temperatures.

Finally, we present table \ref{tab3} to compare the results of
critical temperature from analytical Sturm-Liouville method by
using iterative procedure with numerical values which we
established numerically by using shooting method as explained in
previous section. We take different values of $\alpha$ and
$\kappa$ in this table for three values of $q$ as example.
\begin{center}
\begin{table}[ht]
\begin{tabular}{|c|c|c|c|c|c|c|c|c|}
\hline
&\multicolumn{2}{c|}{$\kappa=0.05 \ , \ \alpha=0.10$}
&\multicolumn{2}{c|}{$\kappa=0.05  \ , \ \alpha=-0.01 $}
&\multicolumn{2}{c|} {$\kappa=0.10  \ , \alpha=-0.05$}
 \\
\hline
$q$ & Analytical & Numerical &Analytical & Numerical
&Analytical & Numerical  \\
\hline
$3/4$   & 0.2622 $\rho^{1/3}$ & 0.2623 $\rho^{1/3}$ &0.2666
$\rho^{1/3}$ & 0.2667 $\rho^{1/3}$ & 0.2639 $\rho^{1/3}$&0.2642
$\rho^{1/3}$ \\
\hline
$1$  &0.1868 $\rho^{1/3}$&0.1882 $\rho^{1/3}$&0.1940$\rho^{1/3}$&0.1959
$\rho^{1/3}$&0.1879 $\rho^{1/3}$&0.1908 $\rho^{1/3}$ \\
\hline
$5/4$  &0.1134 $\rho^{1/3}$&0.1168 $\rho^{1/3}$&0.1208
$\rho^{1/3}$&0.1250 $\rho^{1/3}$&0.1124 $\rho^{1/3}$&0.1177 $\rho^{1/3}$\\
\hline
\end{tabular}
\caption{Comparison of analytical and numerical values of critical
temperature for $\tilde{m}^2=-3$ for certain values of $\kappa$
and $\alpha$.}\label{tab3}
\end{table}
\end{center}
\section{Critical exponent} \label{CriE}
In this section, we propose to analytically calculate the
critical exponent of the Gauss-Bonnet holographic superconductor
with backreaction in the general Power-Maxwell electrodynamics
case for all allowed values of $q$. While we are near the critical
point, $<\mathcal{O}_i>$ is small enough, thus we substitute
Eq.~(\ref{psiF}) into the Eq.~(\ref{eomphiz}) and by using the
fact that in the expansion of $\chi$ Eq.~ (\ref{chiexpand}) the first
term is proportional to $<\mathcal{O}_i>^2$, while we are near the
critical temperature we neglect $\chi'(z)$ and  arrive at
\begin{eqnarray}\label{critphi}
\phi'' -\left(\frac{5-4q}{2q-1}\right) \frac{1}{z} \phi'
-\frac{2r_{+}^{2q-2\Delta_{i}-2} z^{2\Delta_{i}-4q}F^2
\phi'^{2-2q} \phi <\mathcal{O}_i>^2}{(-1)^{2q} q (2q-1)
g(z)}=0,
\end{eqnarray}
where $g(z)$ is defined as in Eq.~(\ref{gz}). Near the critical point,
$T_c \approx T_{0}$, and inspired by Eq.~(\ref{phiPM}), we assume that
Eq. (\ref{critphi}) has the following solution
\begin{eqnarray}\label{phisol}
\phi(z)= AT_{c}(1-z^{\frac{4-2q}{2q-1}})-(A
T_{c})^{3-2q}\left(\frac
{r_{+}^{2q-2\Delta_{i}-2}<\mathcal{O}_{i}>^2}{(-1)^{2q}
q(2q-1)} \right)\Xi (z),
\end{eqnarray}
where
\begin{eqnarray}
A=\frac{4\pi\zeta_{\rm min}}{4-\frac{(4-2q)^{2q}}{3
(2q-1)^{2q-1}}\kappa^2 \zeta_{\rm min}^{2q}}.
\end{eqnarray}
Substituting Eq. (\ref{phisol}) into (\ref{critphi}) and keeping
terms up to $<\mathcal{O}_i>^2$, we reach
\begin{eqnarray}\label{chi}
\Xi''-\left(\frac{5-4q}{2q-1}\right)\frac{\Xi'}{z}-
\frac{(\frac{2q-4}{2q-1})^{2-2q}z^{\eta}(1-z^{\frac{4-2q}{2q-1}})F(z)^2}{g(z)
}=0,
\end{eqnarray}
where
\begin{eqnarray}
\eta=2\Delta_i-4q+\left(\frac{5-4q}{2q-1}\right)(2-2q).
\end{eqnarray}
This is a differential equation for $\Xi(z)$ independent of $r_+$,
$r_{+_c}$ and $<\mathcal{O}_i>$.  Therefore $\Xi(z)$ in any $z$
has a value independent of $T$, $T_{c}$ and order parameter
$<\mathcal{O}_i>$.

The boundary condition for $\phi$ given by Eq. (\ref{BC}), in the
$z$ coordinate, can be rewritten as
\begin{eqnarray}\label{phiz22}
\phi(z) = \mu\left(1 - \frac{\rho^{\frac{1}{2q-1}}}{\mu
r_+^{\frac{4-2q}{2q-1} }}z^{\frac{4-2q}{2q-1} }\right),
\end{eqnarray}
It is reliable while $z \approx 0$, independent of
temperature and order parameter. Also near the
 critical temperature where $\psi$ is small,
  Eq.~(\ref{phiPM}) may be expressed as
\begin{eqnarray}
\phi_0(z)=\frac{\rho^{\frac{1}{2q-1}}}{r_+^{\frac{3}{2q-1}}} r_{+} \left(1-z^{\frac{4-2q}{2 q-1}}\right),
\end{eqnarray}
Since it is valid for all values of $z$, we can
equate the above expression with Eq.
 (\ref{phiz22}) for
$z\rightarrow 0$ to find
\begin{eqnarray}\label{mu} \mu =
\frac{\rho^{\frac{1}{2q-1}}}{r_+^{\frac{3}{2q-1}-1}},
\end{eqnarray}
Since Eq.~(\ref{phiz22}) implies that at infinite boundary $z=0$,
the gauge field is equal to chemical potential, i.e.,
$\phi(z=0)=\mu$. From Eqs. (\ref{Temp}) and (\ref{Hf}), we realize
that $r_+ \propto T$ and it is obvious from Eq.~(\ref{TPM}) that $\rho
\propto T_c^3$. Thus by using Eq.~(\ref{mu}), one can find
\begin{eqnarray}\label{phizero}
\phi(z=0)=\mu=A
\frac{T_{c}^{\frac{3}{2q-1}}}{T^{\frac{4-2q}{2q-1}}}.
\end{eqnarray}
Eq.~(\ref{phisol}) at $z=0$ is equal to it's infinite boundary
value given in Eq.~(\ref{phizero}). Equating Eqs. (\ref{phisol}) and
(\ref{phizero}), we find
\begin{eqnarray}\label{eqtc}
AT_{c}-A \frac{T_{c}^{\frac{3}{2q-1}}}{T^{\frac{4-2q}{2q-1}}}=
(A T_{c})^{3-2q}\left(\frac
{r_{+}^{2q-2\Delta_i-2}<\mathcal{O}_i>^2}{(-1)^{2q}
q(2q-1)} \right)\Xi (0),
\end{eqnarray}
where $\Xi(0)$ is just a constant which can be calculated
numerically from Eq.~(\ref{chi}) with boundary conditions
$\Xi(1)=\Xi'(1)=1$. Using Eqs. (\ref{Temp}) and (\ref{Hf}) for
replacing $r_+$ with $T$ in Eq. (\ref{eqtc}) and then solving the
resulting equation for $<\mathcal{O}_i>$, we get
\begin{eqnarray}
<\mathcal{O}_i>=\gamma T_c^{\Delta_i}\left(\frac{T}{T_c}
\right)^{\Delta_i-q+1}\sqrt{\left(\frac{T_c}{T}\right)^
{\frac{4-2q}{2q-1}}
\left[1-\left(\frac{T}{T_c}\right)^{\frac{4-2q}{2q-1}}\right]},
\end{eqnarray}
where $\gamma$ is a constant independent of $T$ and $T_c$. Using
the fact that $T \approx T_c$, we can rewrite $<\mathcal{O}_i>$ as
\begin{eqnarray}\label{critexp}
<\mathcal{O}_i> \approx \gamma T_c^{\Delta_i}
\sqrt{1-\left(\frac{T}{T_{c}}\right)^{\frac{4-2q}{2q-1}}} \approx \gamma
T_c^{\Delta_i}
\sqrt{1-\left[1-\left(\frac{4-2q}{2q-1}\right)t\right]}\approx
\gamma T_c^{\Delta_i}\sqrt{\left(\frac{4-2q}{2q-1}\right)t} , \nonumber \\
\end{eqnarray}
where $t={(T_c-T)}/{T_c}$. Eq.~(\ref{critexp}) indicates that the
critical exponent $\beta$ of the order parameter is $1/2$
and this result is valid both for  $<\mathcal{O}_{-}>$ and
$<\mathcal{O}_{+}>$. It is obvious that in the presence of
backreaction this exponent for Gauss-Bonnet gravity with
Power-Maxwell field remains unchanged which seems to be a
universal exponent. Let us note that for $q=2$, the expectation
value of the condensation operator vanishes, which means there is
no phase transition in upper bound of $q$.
\section{Conclusion and discussion} \label{Con}
Analytically and based on Sturm-Liouville eigenvalue problem, we
have investigated the properties of $(3+1)$-dimensional $s$-wave
holographic superconductors in the background of five dimensional
Gauss-Bonnet-AdS black holes with Power-Maxwell electrodynamics.
We have considered the case in which the gauge and scalar fields back react on the background geometry. We find out the relation
between critical temperature of phase transition and charge
density is still  $T_c \propto \rho^{1/3}$.  Using the analytical
Sturm-Liouvill method, we have calculated the proportional
constant between the critical temperature and the charge density
for all allowed values of the power parameter $q$, different
values of the Gauss-Bonnet coupling constant $\alpha$, and
backreaction parameter $\kappa$. We realized that decreasing $q$
from Maxwell case ($q=1$) to it's lower bound $(q=1/2)$ increases
the critical temperature, regardless of the values of $\alpha$ and
$\kappa$. Besides, for a fixed values of $q$  and $\kappa$,
critical temperature increases with decreasing the Gauss-Bonnet
coefficient $\alpha$. This means that, increasing $q$ and $\alpha$
will decrease the critical condensation of the scalar field and
make it harder to form. Also, we observed that taking
backreaction into account, decreases the critical temperature
regardless of the values of the other parameters. We have
confirmed these analytical results by providing the numerical
calculations based on the shooting method. Finally, our
investigation of critical exponent indicates that the critical
exponent $\beta$ of the superconducting phase transition for the
five dimensional Power-Maxwell holographic superconductor with
backreaction has the mean field value $1/2$ which seems to
be a universal constant.

\acknowledgments{We thank Shiraz University Research Council. The
work of A.S has been supported financially by Research Institute
for Astronomy and Astrophysics of Maragha (RIAAM), Iran.}

\end{document}